# Effect of molar volume of diffusing elements and cross terms of Onsager formalism (vacancy wind effect) on estimated diffusion coefficients in ternary and multicomponent solid solutions

Aloke Paul*

Department of Materials Engineering, Indian Institute of Science, Bengaluru, India-560012

*Corresponding author, Email: aloke@iisc.ac.in

**Abstract**

The correlation between intrinsic and tracer diffusion coefficients considering the molar volume of the diffusing elements and cross phenomenological constants of Onsager (contributing to the vacancy wind effect) is established in guidance to Manning's analysis. Manning established this correlation assuming constant molar volume. A recent study in binary systems indicated a significant error in estimated data for consideration of constant molar volume if the molar volumes of the diffusing elements are significantly different. It also indicated that considering ideal molar volume variation is essential if a system's non-ideal molar volume variation is unknown. This could not have been practiced until now in ternary and multicomponent systems because of the unavailability of this correlation. This equation scheme is established in this study for conventional diffusion couples (all elements produce diffusion profiles) and constrained, such as pseudo-binary and pseudo-ternary diffusion couples (in which only two or three elements produce the diffusion profiles, respectively). The difference in data calculated considering a constant molar volume and neglecting the vacancy wind effect (which is mainly practiced until today) is compared with the data considering the molar volume of diffusing elements and vacancy wind effect in NiCoFeCr and an imaginary quaternary multi-principal element system. This indicates a significant error in estimated data when considering a constant molar volume if the molar volumes of the diffusing elements are reasonably different. Therefore, the outcome of this study is expected to bring a significant change in diffusion analysis in ternary and multicomponent systems.

## 1. Introduction

The framework for estimating diffusion coefficients following the diffusion couple method in ternary and multicomponent systems was established by Kirkaldy [1] extending the Matano-Boltzmann method proposed for binary systems [2, 3]. Frequently, this is referred to as the Matano-Kirkaldy method. By circumventing the problem of estimating the interdiffusion flux following the Matano method, which needs to identify first the Matano or initial contact plane of the diffusion couple, Whittle and Green [4] proposed to extend the relations established by Sauer-Freise [5], Wagner [6] or Den Broder [7], which do not require locating this plane [8]. Van Loo [9] similarly proposed the calculation of intrinsic flux at the Kirkendall marker plane, which does not need to locate the Matano plane compared to the method proposed earlier by Huemann [10] in guidance to the concept proposed by Matano for interdiffusion coefficients [3]. The same relation was later derived differently by extending the Wagner method by the author of this article [8]. The estimation method of intrinsic flux can also be extended to ternary and multicomponent systems, similar to the calculation of the interdiffusion flux. However, these fluxes are related differently to the diffusion coefficients in ternary/multicomponent systems. We have only one interdiffusion coefficient and two intrinsic diffusion coefficients in a binary system compared to $(n-1)^2$ interdiffusion coefficients and $n(n-1)$ intrinsic diffusion coefficients in ternary and multicomponent systems because of more complex diffusional interactions originating from the thermodynamic correlations [1, 11, 12].

The interdiffusion and intrinsic diffusion coefficients could be correlated earlier considering actual (ideal or non-ideal) molar volume variation [12]. Manning established the correlation between intrinsic and tracer diffusion coefficients considering the vacancy wind effect i.e. contribution from cross phenomenological constants of Onsager formalism [13, 14] but for constant molar volume [15, 16], mentioning the need for correction when molar volume variation is not constant [15]. The actual molar volume variation is never exactly constant since the lattice parameters of two elements are never the same. In real systems, this may vary ideally or non-ideally [17]. Therefore, the current author recently established this correlation for a binary system, considering the actual molar volume filling the missing link [18]. The study in the binary system indicated the importance of considering ideal molar volume variation calculated from the known molar volume of the diffusion elements when the actual molar volume variation (if non-ideal) is unknown.

The molar volume in ternary and multicomponent systems may also vary (almost) ideally or non-ideally with positive/negative deviation from the ideality. The non-availability of intrinsic and tracer diffusion coefficients correlation for ternary and multicomponent systems, considering the cross-phenomenological constants of Onsager for ideal and actual (ideal/non-ideal) molar volume variation, is still an existing problem that needs to be solved. The data are calculated invariably considering constant molar volume variation since the actual molar volume variation, if non-ideal, is unknown. However, in that case, we should be able to calculate more accurate data by considering the ideal molar volume variation, which can be calculated from the known molar volumes of the pure elements. Therefore, the correlation between intrinsic and tracer diffusion coefficients considering cross-phenomenological constants and actual (ideal/non-ideal) molar volume in guidance to Manning's analysis is derived first in this article for ternary and multicomponent conventional diffusion couples in which all the elements produce diffusion profiles. The same is also derived for pseudo-binary and pseudo-ternary diffusion couple [19-22] (methods established recently) in which two or three elements, respectively, develop the diffusion profiles. The difference in calculated intrinsic and interdiffusion coefficients considering constant and ideal molar volume variation from known tracer diffusion coefficients is demonstrated by considering experimentally estimated tracer diffusion coefficients in the NiCoFeCr system and an imaginary system with possible higher molar volume variation and higher difference in tracer diffusivities of elements. The outcome of this study highlights the need to adopt this estimation strategy considering the ideal molar volume variation instead of constant molar volume. One can use this same correlation if the system has non-ideal molar volume variation, and these details will be made available in future. This study is expected to majorly impact overhauling the diffusion analysis in ternary and multicomponent systems.

## 2. Conventional ternary and multicomponent diffusion couples

This section will explain the correlation between different types of diffusion coefficients (tracer, intrinsic and interdiffusion coefficients) in conventional diffusion couples in which all elements produce interdiffusion profiles. The correlations for constrained diffusion couples, such as pseudo-binary (PB) and pseudo-ternary (PT), are established in the next section.

### 2.1 Interdiffusion and intrinsic diffusion coefficients correlation considering the molar volume of elements

Before establishing the correlations between intrinsic and tracer diffusion coefficients, the correlation between interdiffusion and intrinsic diffusion coefficients should be understood considering actual (ideal or non-ideal) molar volume variation. Interdiffusion flux ($\tilde{J}_i$), interdiffusion coefficients $\left(\widetilde{D}_{ij}\right)$ and concentration gradient $\left(\frac{\partial C_j}{\partial x}\right)$ in n-component ternary and multicomponent systems for element *i* without the dependent variable are related by [1, 11, 23, 24]

$$\tilde{J}_i = -\sum_{j=1}^{n} \widetilde{D}_{ij} \frac{\partial C_j}{\partial x} \qquad (1)$$

From standard thermodynamic relations [12], we have

$$\sum_{i=1}^{n} \bar{V}_i \partial C_i = 0 \qquad (2)$$

Where $\bar{V}_i$ is the partial molar volume of element *i*.

Replacing $\partial C_n \left(= -\sum_{i=1}^{n-1} \frac{\bar{V}_i}{\bar{V}_n} \partial C_i\right)$ from Eq. 2 in Eq. 1 for expressing the correlation considering element *n* as the dependent variable, we have

$$\tilde{J}_i = -\sum_{j=1}^{n-1} \widetilde{D}_{ij}^{n} \frac{\partial C_j}{\partial x} \qquad (3a)$$

Therefore, each interdiffusion flux is related to (n-1) independent interdiffusion coefficients, where $\widetilde{D}_{ij}^{n} = \widetilde{D}_{ij} - \frac{\bar{V}_j}{\bar{V}_n} \widetilde{D}_{in}$. For constant molar volume variation (i.e. assuming the partial molar volume of elements $\bar{V}_1 = \bar{V}_1 \dots . = \bar{V}_n$ equal to an average molar volume $V_m$), this is related by $\widetilde{D}_{ij}^{n} = \widetilde{D}_{ij} - \widetilde{D}_{in}$.

The interdiffusion flux can be directly calculated from the concentration profile of a diffusion couple following [4, 7]

$$\tilde{J}_i = -\frac{C_i^+ - C_i^-}{2t}\left[(1 - Y_C^*)\int_{x^{-\infty}}^{x^*} Y_C dx + Y_C^* \int_{x^*}^{x^{+\infty}} (1 - Y_C)\, dx\right] \tag{3b}$$

Where $t$ is the annealing time of the diffusion couple, $Y_C = \frac{C_i - C_i^-}{C_i^+ - C_i^-}$ is the concentration normalized variable, "*" represents the composition or position of interest, $C_i^-$ and $C_i^+$ are the concentration of left- and right-hand side of the diffusion couple, t is the annealing time at a particular temperature of interest [7]. It should be noted here that the interdiffusion fluxes of different elements are related by [12]

$$\sum_{i=1}^{n} \bar{V}_i \tilde{J}_i = 0 \tag{4}$$

Therefore, considering element *n* as the dependent variable, the interdiffusion flux of element *n* is related to the interdiffusion fluxes of other (*n*-1) elements such that we have (n-1) independent interdiffusion fluxes related to total $(n-1)^2$ interdiffusion coefficients.

Similarly, the intrinsic flux of element *i* without the dependent variable is related by [1, 11, 12]

$$J_i = -\sum_{j=1}^{n} D_{ij} \frac{\partial C_j}{\partial x} \tag{5}$$

Again, considering element *n* as the dependent variable and replacing Eq. 2 in Eq. 5, we have

$$J_i = -\sum_{j=1}^{n-1} D_{ij}^n \frac{\partial C_j}{\partial x} \tag{6a}$$

where $D_{ij}^n = D_{ij} - \frac{\bar{V}_j}{\bar{V}_n} D_{in}$. Therefore, each intrinsic flux is related to (n-1) intrinsic diffusion coefficients. Since we have *n* intrinsic fluxes, there are total *n*(*n*-1) intrinsic diffusion coefficients.

The intrinsic fluxes at the Kirkendall marker plane position $(x_m)$ can be calculated from [8, 9, 25]

$$J_i = -\frac{1}{2t}\left[C_i^+ \int_{x^{-\infty}}^{x_m} Y_C dx - C_i^- \int_{x_m}^{x^{+\infty}} (1 - Y_C)\, dx\right] \tag{6b}$$

Following Darken, the interdiffusion and intrinsic fluxes of an element *i* are related to marker velocity ($v_k$) in a diffusion couple by [26]

$$\tilde{J}_i = J_i + v_m C_i \tag{7}$$

We need to replace the marker velocity with intrinsic fluxes. Replacing Eq. 7 in Eq. 4, we have

$$\sum_{i=1}^{n}(\bar{V}_i J_i + v_m \bar{V}_i C_i) = 0 \tag{8}$$

Since $\sum_{i=1}^{n} \bar{V}_i C_i = 1$, we have

$$v_m = -\sum_{i=1}^{n} \bar{V}_i J_i \tag{9}$$

Replacing Eq. 9 in Eq. 7, we have

$$\tilde{J}_i = J_i - C_i \sum_{k=1}^{n} \bar{V}_k J_k \tag{10a}$$

Replacing Eq. 3a and 6a in Eq. 10a, we have

$$\tilde{D}_{ij}^{n} = D_{ij}^{n} - C_i\left(\sum_{k=1}^{n} \bar{V}_k \, D_{kj}^{n}\right) \tag{10b}$$

For constant molar volume (such that $\bar{V}_i = \bar{V}_j = \cdots = \bar{V}_n = V_m$), we have

$$\tilde{D}_{ij}^{n} = D_{ij}^{n} - N_i\left(\sum_{k=1}^{n} D_{kj}^{n}\right) \tag{10c}$$

which has been used until now.

### 2.2 Correlation between intrinsic and tracer diffusion coefficients

Manning [16] proposed the relation between the intrinsic flux and tracer diffusion coefficients considering Onsager's formalism (Eq. 1) as

$$J_i = -\frac{C_i D_i^*}{RT}\frac{\partial \mu_i}{dx} - \beta N_i D_i^* J_v \tag{11a}$$

where $\beta = \frac{2}{(M_o+2)\sum_{j=1}^{n} V_j C_j D_j^*}$ when molar volume varies with composition [18] and $M_o$ is the structure factor. In FCC crystal, this is 7.15 [15]. $J_v$ is the flux of vacancies. This is established in correlation to the Onsager formalism [13, 14]

$$J_i = -L_{i1}\frac{\partial \mu_1}{\partial x} \ldots - L_{ii}\frac{\partial \mu_i}{\partial x} \ldots - L_{in}\frac{\partial \mu_n}{\partial x} \tag{11b}$$

where $L_{ii}$ is the main phenomenological constant related to the chemical potential gradient of the same element $\left(\frac{\partial \mu_i}{\partial x}\right)$, and $L_{ij}$ is the cross phenomenological constant of the element related to the chemical potential of another element *j*, i.e. $\left(\frac{\partial \mu_j}{\partial x}\right)$.

The first term of Eq. 11a is related to Onager's main phenomenological constant $(L_{ii})$, and the second term is related to all the cross terms $(L_{ij})$.

We need to replace the vacancy flux term with tracer diffusion coefficients. The vacancy flux is related to the intrinsic fluxes of elements by [12, 27]

$$V_m J_v = -\sum_{i=1}^{n} V_i J_i \tag{12}$$

Replacing Eq. 11 in Eq. 12, we have

$$V_m J_v = \bar{V}_1 \frac{C_1 D_1^*}{RT}\frac{\partial \mu_1}{dx} + \bar{V}_1 \beta N_1 D_1^* J_v + \bar{V}_2 \frac{C_2 D_2^*}{RT}\frac{\partial \mu_2}{dx} + \bar{V}_2 \beta N_2 D_2^* J_v + \cdots + \bar{V}_n \frac{C_n D_n^*}{RT}\frac{\partial \mu_n}{dx} + \bar{V}_n \beta N_n D_n^* J_v$$

$$(V_m - \beta \bar{V}_1 N_1 D_1^* - \beta \bar{V}_2 N_2 D_2^* \ldots - \beta \bar{V}_n N_n D_n^*) J_v = \bar{V}_1 \frac{C_1 D_1^*}{RT}\frac{\partial \mu_1}{dx} + \bar{V}_2 \frac{C_2 D_2^*}{RT}\frac{\partial \mu_2}{dx} \ldots + \bar{V}_n \frac{C_n D_n^*}{RT}\frac{\partial \mu_n}{dx}$$

$$J_v = \frac{1}{V_m - \beta \sum_{j=1}^{n} \bar{V}_j N_j D_j^*} \sum_{j=1}^{n} \frac{\bar{V}_j C_j D_j^*}{RT}\frac{\partial \mu_j}{dx} \tag{13}$$

Replacing Eq. 13 in Eq. 11, we have

$$J_i = -\frac{C_i D_i^*}{RT}\frac{\partial \mu_i}{dx} - N_i D_i^* \frac{\beta}{V_m - \beta \sum_{j=1}^{n} \bar{V}_j N_j D_j^*} \sum_{j=1}^{n} \frac{\bar{V}_j C_j D_j^*}{RT}\frac{\partial \mu_j}{dx}$$

$$J_i = -\frac{C_i D_i^*}{RT}\frac{\partial \mu_i}{dx} - C_i D_i^* \frac{\beta}{1 - \beta \sum_{j=1}^{n} \bar{V}_j C_j D_j^*} \sum_{j=1}^{n} \frac{\bar{V}_j C_j D_j^*}{RT}\frac{\partial \mu_j}{dx}$$

$$J_i = -\frac{C_i D_i^*}{RT}\frac{\partial \mu_i}{dx} - (\xi_a C_i D_i^*) \sum_{j=1}^{n} \frac{\bar{V}_j C_j D_j^*}{RT}\frac{\partial \mu_j}{dx} \tag{14}$$

where $\xi_a = \frac{\beta}{1 - \beta \sum_{j=1}^{n} \bar{V}_j C_j D_j^*} = \frac{\frac{2}{(M_O + 2)\sum_{j=1}^{n} \bar{V}_j C_j D_j^*}}{1 - \frac{2\sum_{j=1}^{n} \bar{V}_j C_j D_j^*}{(M_O + 2)\sum_{j=1}^{n} \bar{V}_j C_j D_j^*}} = \frac{2}{M_O \sum_{j=1}^{n} \bar{V}_j C_j D_j^*}$

This can be expanded for element *i* as

$$J_i = -\frac{C_i D_i^*}{RT}\frac{\partial \mu_i}{dx} - (\xi_a C_i D_i^*) \frac{\bar{V}_1 C_1 D_1^*}{RT}\frac{\partial \mu_1}{dx} \ldots - (\xi_a C_i D_i^*) \frac{\bar{V}_i C_i D_i^*}{RT}\frac{\partial \mu_i}{dx} \ldots - (\xi_a C_i D_i^*) \frac{\bar{V}_n C_n D_n^*}{RT}\frac{\partial \mu_n}{dx} \tag{15a}$$

This molar flux can be expressed as volume flux by

$$\bar{V}_i J_i = -\frac{\bar{V}_i C_i D_i^*}{RT}\frac{\partial \mu_i}{dx} - (\xi_a \bar{V}_i C_i D_i^*) \frac{\bar{V}_1 C_1 D_1^*}{RT}\frac{\partial \mu_1}{dx} \ldots - (\xi_a \bar{V}_i C_i D_i^*) \frac{\bar{V}_i C_i D_i^*}{RT}\frac{\partial \mu_i}{dx} \ldots - (\xi_a \bar{V}_i C_i D_i^*) \frac{\bar{V}_n C_n D_n^*}{RT}\frac{\partial \mu_n}{dx} \tag{15b}$$

Similarly, the Onsager formalism [13, 14] for element *i* as volume flux for constant pressure and temperature, can be expressed from Eq. 11b as

$$\bar{V}_i J_i = -\bar{V}_i L_{i1} \frac{\partial \mu_1}{\partial x} \ldots - \bar{V}_i L_{ii} \frac{\partial \mu_i}{\partial x} \ldots - \bar{V}_i L_{in} \frac{\partial \mu_n}{\partial x} \quad (16)$$

Comparing Eq. 15b and 16, the Onsager phenomenological constants can be expressed with diffusion coefficients as

$$L_{ii} = -\frac{C_i D_i^*}{RT}\left(1 + \xi_a \bar{V}_i C_i D_i^*\right) \quad (17a)$$

$$\bar{V}_i L_{ij} = \bar{V}_j L_{ji} = \xi_a \frac{\bar{V}_i C_i D_i^* \bar{V}_j C_j D_j^*}{RT} \quad (17b)$$

For the sake of comparison to Fick's law of diffusion relating intrinsic and tracer diffusion coefficients, the chemical potential gradient must be converted to the concentration gradient. The chemical potential gradient of element *i* can be expressed as

$$\frac{\partial \mu_i}{\partial x} = \frac{\partial \mu_i}{\partial C_1}\frac{\partial C_1}{\partial x} \ldots + \frac{\partial \mu_i}{\partial C_i}\frac{\partial C_i}{\partial x} \ldots + \frac{\partial \mu_i}{\partial C_{(n-1)}}\frac{\partial C_{(n-1)}}{\partial x} + \frac{\partial \mu_i}{\partial C_n}\frac{\partial C_n}{\partial x} \quad (18)$$

The chemical potential and activity $(a_i)$ of an element are related by $\mu_i = \mu_i^o + RTlna_i$. Therefore, Eq. 17 can be expressed as

$$\frac{\partial \mu_i}{\partial x} = \frac{RT\partial lna_i}{\partial C_1}\frac{\partial C_1}{\partial x} \ldots + \frac{\partial \mu_i}{\partial C_i}\frac{\partial C_i}{\partial x} \ldots + \frac{RT\partial lna_i}{\partial C_{(n-1)}}\frac{\partial C_{(n-1)}}{\partial x} + \frac{RT\partial lna_i}{\partial C_n}\frac{\partial C_n}{\partial x}$$

$$\frac{\partial \mu_i}{\partial x} = \frac{RT\partial lna_i}{C_1 \partial lnC_1}\frac{\partial C_1}{\partial x} \ldots + \frac{RT\partial lna_i}{C_i \partial lnC_i}\frac{\partial C_i}{\partial x} \ldots + \frac{RT\partial lna_i}{C_{(n-1)}\partial lnC_{(n-1)}}\frac{\partial C_{(n-1)}}{\partial x} + \frac{RT\partial lna_i}{C_n \partial lnC_n}\frac{\partial C_n}{\partial x} \quad (19)$$

Replacing $\partial C_n \left(= -\sum_{i=1}^{n-1}\frac{\bar{V}_i}{\bar{V}_n}\partial C_i\right)$ from Eq. 2, Eq. 19 can be rewritten as

$$\frac{\partial \mu_i}{\partial x} = \frac{RT}{C_1}\left(\frac{\partial lna_i}{\partial lnC_1} - \frac{V_1 C_1}{V_n C_n}\frac{\partial lna_i}{\partial lnC_n}\right)\frac{\partial C_1}{\partial x} \ldots + \frac{RT}{C_i}\left(\frac{\partial lna_i}{\partial lnC_i} - \frac{V_i C_i}{V_n C_n}\frac{\partial lna_i}{\partial lnC_n}\right)\frac{\partial C_i}{\partial x} + \ldots + \frac{RT}{C_{n-1}}\left(\frac{\partial lna_i}{\partial lnC_{n-1}} - \frac{V_{n-1}C_{n-1}}{V_n C_n}\frac{\partial lna_i}{\partial lnC_n}\right)\frac{\partial C_{n-1}}{\partial x}$$

$$\frac{\partial \mu_i}{\partial x} = \frac{RT}{C_1}\left(\varphi_{i1} - \frac{V_1 C_1}{V_n C_n}\varphi_{in}\right)\frac{\partial C_1}{\partial x} \ldots . \frac{RT}{C_i}\left(\varphi_{ii} - \frac{V_i C_i}{V_n C_n}\varphi_{in}\right)\frac{\partial C_i}{\partial x} \ldots .. + \frac{RT}{C_{n-1}}\left(\varphi_{in-1} - \frac{C_{n-1}}{N_n}\varphi_{in}\right)\frac{\partial C_{n-1}}{\partial x}$$

$$\frac{\partial \mu_i}{\partial x} = \frac{RT}{C_1}\ \varphi_{i1}^n \frac{\partial C_1}{\partial x} \ldots + \frac{RT}{C_i}\varphi_{ii}^n \frac{\partial C_i}{\partial x} \ldots + \frac{RT}{C_{n-1}}\varphi_{i(n-1)}^n \frac{\partial C_{n-1}}{\partial x}$$

$$\frac{\partial \mu_i}{\partial x} = \sum_{j=1}^{n-1}\frac{RT}{C_j}\ \varphi_{ij}^n \frac{\partial C_j}{\partial x} \quad (20a)$$

where the thermodynamic factor considering component *n* as the dependent variable ($\varphi_{ij}^{n}$) and considering the actual molar volume of the elements (instead of constant molar volume variation) is expressed as

$\varphi_{ij}^{n} = \left(\varphi_{ij} - \frac{V_j C_j}{V_n C_n}\varphi_{in}\right) = \frac{\partial ln a_i}{\partial ln C_j} - \frac{V_j C_j}{V_n C_n}\frac{\partial ln a_i}{\partial ln C_n}$ . For constant molar volume, this reduces to $\emptyset_{ij}^{n} = \left(\emptyset_{ij} - \frac{N_j}{N_n}\emptyset_{in}\right) = \frac{\partial ln a_i}{\partial ln N_j} - \frac{N_j}{N_n}\frac{\partial ln a_i}{\partial ln N_n}$, the parameter used in the equation scheme established earlier by Manning considering the constant molar volume [16].

Replacing Eq. 20a in Eq. 15a, we have

$$J_i = -\frac{C_i D_i^*}{RT}\sum_{j=1}^{n-1}\frac{RT}{C_j}\ \varphi_{ij}^{n}\frac{\partial C_j}{\partial x} - (\xi_a C_i D_i^*)\frac{\overline{V}_1 C_1 D_1^*}{RT}\sum_{j=1}^{n-1}\frac{RT}{C_j}\ \varphi_{1j}^{n}\frac{\partial C_j}{\partial x}\dots - (\xi_a C_i D_i^*)\frac{\overline{V}_i C_i D_i^*}{RT}\sum_{j=1}^{n-1}\frac{RT}{C_j}\ \varphi_{ij}^{n}\frac{\partial C_j}{\partial x}\dots - (\xi_a C_i D_i^*)\frac{\overline{V}_n C_n D_n^*}{RT}\sum_{j=1}^{n-1}\frac{RT}{C_j}\ \varphi_{nj}^{n}\frac{\partial C_j}{\partial x} \qquad (21)$$

Similarly, Eq. 6 expressing the intrinsic flux of element *i* can be expanded to

$$J_i = -D_{i1}^{n}\frac{\partial C_1}{\partial x}\dots - D_{ij}^{n}\frac{\partial C_j}{\partial x}\dots . -D_{i(n-1)}^{n-1}\frac{\partial C_{n-1}}{\partial x} \qquad (22)$$

Comparing Eq. 21 and 22, the intrinsic diffusion coefficient can be expressed as

$$D_{ij}^{n} = \frac{C_i}{C_j}D_i^*\varphi_{ij}^{n} + \left(\xi_a C_i D_i^*\right)\frac{C_1}{C_j}\overline{V}_1 D_1^*\varphi_{1j}^{n} + \cdots + \left(\xi_a C_i D_i^*\right)\frac{C_i}{C_j}\overline{V}_i D_i^*\varphi_{ij}^{n} + \ldots + \left(\xi_a C_i D_i^*\right)\frac{C_n}{C_j}\overline{V}_n D_n^*\varphi_{nj}^{n}$$

$$D_{ij}^{n} = \frac{C_i}{C_j}D_i^*\varphi_{ij}^{n}\left(1 + W_{ij}^{n,a}\right) \qquad (23)$$

Where the vacancy wind effect for the intrinsic diffusion coefficients for actual molar volume (ideal or non-ideal) is

$$1 + W_{ij}^{n,a} = 1 + \frac{\xi_a}{\varphi_{ij}^{n}}\sum_{k=1}^{n}\overline{V}_k C_k D_k^*\varphi_{kj}^{n} = 1 + \frac{2\sum_{k=1}^{n}\overline{V}_k C_k D_k^*\varphi_{kj}^{n}}{\varphi_{ij}^{n} M_o \sum_{j=1}^{n}\overline{V}_j C_j D_j^*}$$

Note here that for constant molar volume $\overline{V}_i = \overline{V}_j = V_m$, Eq. 23 reduces to the equation derived earlier by Manning considering constant molar volume as [16]

$$D_{ij}^{n} = \frac{N_i}{N_j}D_i^*\emptyset_{ij}^{n}\left(1 + W_{ij}^{n,c}\right) \qquad (24)$$

where $1 + W_{ij}^{n} = 1 + \frac{\xi}{\emptyset_{ij}^{n}}\sum_{k=1}^{n}N_k D_k^*\emptyset_{kj}^{n} = 1 + \frac{2}{M_o\emptyset_{ij}^{n}}\frac{\sum_{k=1}^{n}N_k D_k^*\emptyset_{kj}^{n}}{\sum_{k=1}^{n}N_k D_k^*}$ where $\emptyset_{ij}^{n} = \emptyset_{ij} - \frac{N_j}{n_n}\emptyset_{in} = \frac{\partial ln a_i}{\partial ln N_j} - \frac{N_j}{N_n}\frac{\partial ln a_i}{\partial ln N_n}$. Therefore, the thermodynamic factors ($\emptyset_{ij}^{n}$) are calculated from activity ($a_i$) vs. composition ($N_j$). For considering the actual molar volume variation (ideal/non-idea), the thermodynamic factor ($\varphi_{ij}^n$) is calculated by taking the gradient of $lna_i$ vs. $lnC_j$ plot (where $C_j = N_j/V_m$) instead of taking the gradient of $lna_i$ vs. $lnN_j$ plot for $\emptyset_{ij}^n$. Please also note that Manning did not express the equations with dependent variables, which are measurable parameters. He expressed the relationship for $D_{ij}$ instead of $D_{ij}^n = D_{ij} - D_{in}$ for constant molar volume. One can express $D_{ij}$ and $D_{in}$ considering the relation he proposed, and then Eq. 24 can be established by taking $D_{ij} - D_{in}$ [16].

### 3. Diffusion coefficient correlations for constrained diffusion couples

Because of the difficulties of intersecting (*n*-1) conventional diffusion profiles in multicomponent space for the estimation of interdiffusion coefficients, the concepts of constrained diffusion couples (i.e. keeping one or more elements constant), such as pseudo-binary (PB) and pseudo-ternary (PT) diffusion couple methods were established, which facilitate the estimation diffusion coefficients of either certain elements systematically [19-21] or all elements by designing the diffusion couples strategically in a combination of constrained and conventional diffusion couples [29-30]. The estimations following these methods have been conducted until now, considering the constant molar volume similar to that of conventional diffusion couples. The aim here is to establish equation schemes that consider the molar volume of elements.

#### 3.1 Pseudo-binary diffusion couple method

Let us consider a n-component system in which elements 2 to (*n*-1) remain constant, and only elements 1 and *n* produce the interdiffusion profiles in a PB diffusion couple. Therefore, the interdiffusion coefficient of elements 1 and *n* can be expressed from Eq. 3 (for $\frac{\partial C_2}{\partial x}, \ldots \frac{\partial C_{n-1}}{\partial x} = 0$) as

$$\tilde{J}_1 = -\tilde{D}_{11}^n \frac{\partial C_1}{\partial x} \quad (25a)$$

$$\tilde{J}_n = -\tilde{D}_{nn}^1 \frac{\partial C_n}{\partial x} \quad (25b)$$

In a PB diffusion couple, we have (from Eq. 2)

$$\bar{V}_1 \partial C_1 + \bar{V}_n \partial C_n = 0 \tag{26}$$

Further, in the PB diffusion couple from Eq. 4, we have

$$\bar{V}_1 \tilde{J}_1 + \bar{V}_n \tilde{J}_n = 0 \tag{27}$$

Replacing Eq. 25 in Eq. 27, we have

$$\bar{V}_1 \widetilde{D}_{11}^{n} \frac{\partial C_1}{\partial x} + \bar{V}_n \widetilde{D}_{nn}^{1} \frac{\partial C_n}{\partial x} = 0 \tag{28}$$

Therefore, considering Eq. 26, we have only one interdiffusion coefficient in a PB diffusion couple (similar to the binary system), such that

$$\widetilde{D}_{11}^{n} = \widetilde{D}_{nn}^{1} = \widetilde{D}_{PB} \tag{29}$$

Similarly, the intrinsic diffusion coefficients can be expressed as

$$J_1 = -D_{11}^{n} \frac{\partial C_1}{\partial x} \tag{30a}$$

$$J_n = -D_{n1}^{n} \frac{\partial C_1}{\partial x} = -D_{nn}^{1} \frac{\partial C_n}{\partial x} \tag{30b}$$

Note that the intrinsic flux can be expressed with element *n* as the dependent variable when related to the concentration gradient of element 1 or 1 as the dependent variable when related to the concentration gradient of element *n* since

$$D_{n1}^{n} \frac{\partial C_1}{\partial x} = -\frac{\bar{V}_n}{\bar{V}_1} D_{n1}^{n} \frac{\partial C_n}{\partial x} = -\frac{\bar{V}_n}{\bar{V}_1}\left(D_{n1} - \frac{\bar{V}_1}{\bar{V}_n} D_{nn}\right)\frac{\partial C_n}{\partial x} = \left(D_{nn} - \frac{\bar{V}_n}{\bar{V}_1} D_{n1}\right)\frac{\partial C_n}{\partial x} = D_{nn}^{1} \frac{\partial C_n}{\partial x}$$

Further, note that elements 1 and n in a pseudo-binary diffusion couple will diffuse in the opposite direction. Therefore, $D_{n1}^{n}$ will be estimated with a negative sign since it is related to the concentration gradient of element 1 instead of element n, although it is related to the intrinsic flux of element n. Even the thermodynamic factor $\varphi_{n1}^{n}$ will have a negative sign that relates to the positive value of the tracer diffusion coefficient of element *n*. One may prefer to express this intrinsic flux with the concentration gradient of element n such that the intrinsic diffusion coefficient $D_{nn}^{1}$ has a positive value. This is similar to the analysis in a binary system, in which the intrinsic flux of an element is expressed with the concentration gradient of the same element.

To establish the correlation between interdiffusion and intrinsic diffusion coefficients in a PB diffusion couple, replacing Eq. 30 in Eq. 8, we have

$$\bar{V}_1 J_1 + \bar{V}_n J_n + v_m(\bar{V}_1 C_1 + \bar{V}_n C_n) = 0$$

$$v_m = -\frac{(\bar{V}_1 J_1 + \bar{V}_n J_n)}{\bar{V}_1 C_1 + \bar{V}_n C_n} \tag{31}$$

Replacing Eq. 31 in Eq. 7, we have

$$\tilde{J}_{i(1,n)} = J_i - \frac{C_i}{\bar{V}_1 C_1 + \bar{V}_n C_n}(\bar{V}_1 J_1 + \bar{V}_n J_n) = J_i - M_i^{PB}(J_1 + J_n) \tag{30}$$

Replacing Eq. 25 and 30 in the above equation, we have (let's say for $\tilde{J}_n$, which will give the same outcome for considering $\tilde{J}_1$)

$$\widetilde{D}_{PB}\frac{\partial C_n}{\partial x} = D_{nn}^1\frac{\partial C_n}{\partial x} - \frac{C_n}{\bar{V}_1 C_1 + \bar{V}_n C_n}\left(\bar{V}_1 D_{11}^n\frac{\partial C_1}{\partial x} + \bar{V}_n D_{nn}^1\frac{\partial C_n}{\partial x}\right)$$

Replacing Eq. 26, we have

$$\widetilde{D}_{PB}\frac{\partial C_n}{\partial x} = D_{nn}^1\frac{\partial C_n}{\partial x} - \frac{C_n}{\bar{V}_1 C_1 + \bar{V}_n C_n}\left(-\bar{V}_n D_{11}^n\frac{\partial C_n}{\partial x} + \bar{V}_n D_{nn}^1\frac{\partial C_n}{\partial x}\right)$$

$$\widetilde{D}_{PB} = \frac{\bar{V}_n C_n}{\bar{V}_1 C_1 + \bar{V}_n C_n} D_{11}^n + \frac{\bar{V}_1 C_1}{\bar{V}_1 C_1 + \bar{V}_n C_n} D_{nn}^1$$

$$\widetilde{D}_{PB} = Z_n^{PB} D_{11}^n + Z_1^{PB} D_{nn}^1 \tag{31a}$$

where $Z_i^{PB} = \frac{\bar{V}_i C_i}{\bar{V}_1 C_1 + \bar{V}_n C_n}$ is a PB-normalized composition. Note here that $\bar{V}_1 C_1 + \bar{V}_n C_n \neq 1$ in the PB system (since other elements are present which do not produce the diffusion profile) compared to $\bar{V}_1 C_1 + \bar{V}_n C_n = 1$ in a binary system of elements 1 and n.

Therefore, this correlation in a binary system is expressed as [12,18]

$$\widetilde{D}_B = \bar{V}_n C_n D_1 + \bar{V}_1 C_1 D_2 \tag{31b}$$

where $\widetilde{D}_B$ is the interdiffusion coefficient $D_1$ and $D_n$ are the intrinsic diffusion coefficients of elements.

Further, in a PB diffusion couple, since only two elements (1 and n) produce the diffusion profiles, Eq. 15a can be expressed as

$$J_{i(1\ and\ n)} = -\frac{C_i D_i^*}{RT}\frac{\partial \mu_i}{dx} - (\xi_a C_i D_i^*)\frac{\bar{V}_1 C_1 D_1^*}{RT}\frac{\partial \mu_1}{dx} - (\xi_a C_i D_i^*)\frac{\bar{V}_n C_n D_n^*}{RT}\frac{\partial \mu_n}{dx} \tag{32}$$

The chemical potential gradient can be expressed from Eq. 20 as

$$\frac{\partial \mu_1}{\partial x} = \frac{RT}{C_1}\ \varphi_{11}^{n} \frac{\partial C_1}{\partial x} \text{ and } \frac{\partial \mu_n}{\partial x} = \frac{RT}{C_n}\ \varphi_{nn}^{1} \frac{\partial C_n}{\partial x} \quad (33)$$

Replacing Eq. 33 in Eq. 32, we have

$$J_1 = -D_1^*\ \varphi_{11}^{n} \frac{\partial C_1}{\partial x} - (\xi_{PB} C_1 D_1^*)\bar{V}_1 D_1^*\ \varphi_{11}^{n} \frac{\partial C_1}{\partial x} - (\xi_{PB} C_1 D_1^*)\bar{V}_n D_n^*\ \varphi_{nn}^{1} \frac{\partial C_n}{\partial x}$$

$$J_n = -D_n^* \varphi_{nn}^{1} \frac{\partial C_n}{\partial x} - (\xi_{PB} C_n D_n^*)\bar{V}_1 D_1^*\ \varphi_{11}^{n} \frac{\partial C_1}{\partial x} - (\xi_{PB} C_n D_n^*)\bar{V}_n D_n^*\ \varphi_{nn}^{1} \frac{\partial C_n}{\partial x}$$

Replacing Eq. 26, we have

$$J_1 = -D_1^*\ \varphi_{11}^{n} \frac{\partial C_1}{\partial x} - (\xi_{PB} C_1 D_1^*)\bar{V}_1 D_1^*\ \varphi_{11}^{n} \frac{\partial C_1}{\partial x} + (\xi_{PB} C_1 D_1^*)\bar{V}_1 D_n^*\ \varphi_{nn}^{1} \frac{\partial C_1}{\partial x}$$

$$J_n = -D_n^* \varphi_{nn}^{1} \frac{\partial C_n}{\partial x} + (\xi_{PB} C_n D_n^*)\bar{V}_n D_1^*\ \varphi_{11}^{n} \frac{\partial C_n}{\partial x} - (\xi_{PB} C_n D_n^*)\bar{V}_n D_n^*\ \varphi_{nn}^{1} \frac{\partial C_n}{\partial x}$$

Comparing the equations above with Eq. 30, we have

$$D_{11}^{n} = D_1^*[\ \varphi_{11}^{n} + \xi_{PB}\bar{V}_1 C_1 (D_1^*\ \varphi_{11}^{n} - D_n^*\ \varphi_{nn}^{1})] = D_1^* \varphi_{11}^{n} (1 + W_{11}^{PB}) \quad (34a)$$

$$D_{nn}^{1} = D_n^*[\varphi_{nn}^{1} - \xi_{PB}\bar{V}_n C_n (D_1^*\ \varphi_{11}^{n} - D_n^*\ \varphi_{11}^{n})] = D_n^* \varphi_{nn}^{1} (1 - W_{nn}^{PB}) \quad (34b)$$

Such that the vacancy wind effects are on elements 1 and n are $(1 + W_{11}^{PB})$ and $(1 - W_{nn}^{PB})$ such that $1 + W_{11}^{PB} = \frac{\xi_{PB}\bar{V}_1 C_1 (D_1^*\ \varphi_{11}^{n} - D_n^*\ \varphi_{nn}^{1})}{\varphi_{11}^{1}}$ and $1 - W_{nn}^{PB} = \frac{\xi_{PB}\bar{V}_n C_n (D_1^* \varphi_{11}^{n} - D_n^*\ \varphi_{nn}^{1})}{\varphi_{nn}^{1}}$.

In this pseudo-binary system of diffusing elements 1 and *n*, we have

$$\xi_{PB} = \frac{2}{M_o \sum_{j=1}^{n} \bar{V}_j C_j D_j^*} = \frac{2}{M_o(\bar{V}_1 C_1 D_1^* + \bar{V}_n C_n D_n^*)} \quad (35)$$

Therefore, the vacancy wind effects in a PB diffusion couple are expressed as

$$1 + W_{11}^{PB} = 1 + \frac{2\bar{V}_1 C_1 (D_1^*\ \varphi_{11}^{n} - D_n^*\ \varphi_{nn}^{1})}{\varphi_{11}^{1} M_o(\bar{V}_1 C_1 D_1^* + \bar{V}_n C_n D_n^*)} \text{ and } W_{nn}^{PB} = 1 - \frac{2\bar{V}_n C_n (D_1^*\ \varphi_{11}^{n} - D_n^*\ \varphi_{nn}^{1})}{\varphi_{11}^{1} M_o(\bar{V}_1 C_1 D_1^* + \bar{V}_n C_n D_n^*)}$$

These are very similar to the correlation for the binary system of elements 1 and *n*, which are expressed ($\xi_{PB} = \xi_B$) as [18]

$$D_1 = D_1^*[\varphi_1 + \xi_B \bar{V}_1 C_1 (D_1^* \varphi_1 - D_n^* \varphi_n)] = D_1^* \varphi_1 (1 + W_1^B) \quad (36b)$$

$$D_n = D_n^*[\varphi_n - \xi_B \bar{V}_n C_n (D_1^* \varphi_1 - D_n^* \varphi_n)] = D_n^* \varphi_n (1 - W_n^B) \quad (36b)$$

Where the vacancy wind effects are

$$1 + W_1^B = 1 + \frac{2\bar{V}_1 C_1 (D_1^* \varphi_1 - D_n^* \varphi_n)}{\varphi_1 M_o(\bar{V}_1 C_1 D_1^* + \bar{V}_n C_n D_n^*)} \text{ and } 1 + W_1^B = 1 - \frac{2\bar{V}_n C_n (D_1^* \varphi_1 - D_n^* \varphi_n)}{\varphi_n M_o(\bar{V}_1 C_1 D_1^* + \bar{V}_n C_n D_n^*)}$$

### 3.2 Pseudo-Ternary Diffusion Couple Method

In a pseudo-ternary system (PT), let's say of elements 1, 2 and n, The interdiffusion coefficients Eq. 3 (for $\frac{\partial C_3}{\partial x}, \dots \frac{\partial C_{n-1}}{\partial x} = 0$) can be expressed as

$$\tilde{J}_1 = -\tilde{D}_{11}^n \frac{\partial C_1}{\partial x} - \tilde{D}_{12}^n \frac{\partial C_2}{\partial x} \quad (37a)$$

$$\tilde{J}_2 = -\tilde{D}_{21}^n \frac{\partial C_1}{\partial x} - \tilde{D}_{22}^n \frac{\partial C_2}{\partial x} \quad (37b)$$

Similarly, the intrinsic diffusion coefficients can be expressed as

$$J_1 = -D_{11}^n \frac{\partial C_1}{\partial x} - D_{12}^n \frac{\partial C_2}{\partial x} \quad (38a)$$

$$J_2 = -D_{21}^n \frac{\partial C_1}{\partial x} - D_{22}^n \frac{\partial C_2}{\partial x} \quad (38b)$$

$$J_n = -D_{n1}^n \frac{\partial C_1}{\partial x} - D_{n2}^n \frac{\partial C_2}{\partial x} \quad (38c)$$

In a PT diffusion couple, replacing Eq. 38 in Eq. 8, we have

$$\bar{V}_1 J_1 + \bar{V}_2 J_2 + \bar{V}_n J_n + v_m(\bar{V}_1 C_1 + \bar{V}_2 C_2 + \bar{V}_n C_n) = 0$$

$$v_m = -\frac{(\bar{V}_1 J_1 + \bar{V}_2 J_2 + \bar{V}_n J_n)}{\bar{V}_1 C_1 + \bar{V}_2 C_2 + \bar{V}_n C_n} \quad (39)$$

Replacing Eq. 39 in Eq. 7, we have

$$\tilde{J}_{i(1,2)} = J_i - \frac{C_i}{\bar{V}_1 C_1 + \bar{V}_2 C_2 + \bar{V}_n C_n} (\bar{V}_1 J_1 + \bar{V}_2 J_2 + \bar{V}_n J_n) \quad (40)$$

Unlike the conventional ternary 1, 2 and n diffusion couple, where all the elements produce the diffusion profiles (such that $\bar{V}_1 C_1 + \bar{V}_2 C_2 + \bar{V}_n C_n = 1$), in a PT diffusion couple, these elements only produce diffusion profiles, keeping other elements constant (such that $\bar{V}_1 C_1 + \bar{V}_2 C_2 + \bar{V}_n C_n \neq 1$. Two independent interdiffusion fluxes can be, therefore, expressed as

$$\tilde{J}_1 = J_1 - \frac{C_1}{\bar{V}_1 C_1 + \bar{V}_2 C_2 + \bar{V}_n C_n} (\bar{V}_1 J_1 + \bar{V}_2 J_2 + \bar{V}_n J_n) \quad (41a)$$

$$\tilde{J}_2 = J_2 - \frac{C_2}{\bar{V}_1 C_1 + \bar{V}_2 C_2 + \bar{V}_n C_n} (\bar{V}_1 J_1 + \bar{V}_2 J_2 + \bar{V}_n J_n) \quad (41b)$$

Replacing Eq. 37 and 38 in Eq. 41a, we have

$$\widetilde{D}_{11}^{n}\frac{\partial C_1}{\partial x}+\widetilde{D}_{12}^{n}\frac{\partial C_2}{\partial x}=D_{11}^{n}\frac{\partial C_1}{\partial x}+D_{12}^{n}\frac{\partial C_2}{\partial x}-\frac{C_1}{\bar{V}_1C_1+\bar{V}_2C_2+\bar{V}_nC_n}\left[\bar{V}_1\left(D_{11}^{n}\frac{\partial C_1}{\partial x}+D_{12}^{n}\frac{\partial C_2}{\partial x}\right)+\bar{V}_2\left(D_{21}^{n}\frac{\partial C_1}{\partial x}+D_{22}^{n}\frac{\partial C_2}{\partial x}\right)+\bar{V}_n\left(D_{n1}^{n}\frac{\partial C_1}{\partial x}+D_{n2}^{n}\frac{\partial C_2}{\partial x}\right)\right] \quad (42a)$$

Similarly, replacing Eq. 37 and 38 in Eq. 41b, we have

$$\widetilde{D}_{21}^{n}\frac{\partial C_1}{\partial x}+\widetilde{D}_{22}^{n}\frac{\partial C_2}{\partial x}=D_{21}^{n}\frac{\partial C_1}{\partial x}+D_{22}^{n}\frac{\partial C_2}{\partial x}-\frac{C_2}{\bar{V}_1C_1+\bar{V}_2C_2+\bar{V}_nC_n}\left[\bar{V}_1\left(D_{11}^{n}\frac{\partial C_1}{\partial x}+D_{12}^{n}\frac{\partial C_2}{\partial x}\right)+\bar{V}_2\left(D_{21}^{n}\frac{\partial C_1}{\partial x}+D_{22}^{n}\frac{\partial C_2}{\partial x}\right)+\bar{V}_n\left(D_{n1}^{n}\frac{\partial C_1}{\partial x}+D_{n2}^{n}\frac{\partial C_2}{\partial x}\right)\right] \quad (42b)$$

Comparing the $\frac{\partial C_1}{\partial x}$ and $\frac{\partial C_2}{\partial x}$ on both sides, we have

$$\widetilde{D}_{11}^{n}=D_{11}^{n}-\frac{\bar{V}_1C_1}{\bar{V}_1C_1+\bar{V}_2C_2+\bar{V}_nC_n}\left[D_{11}^{n}+\frac{\bar{V}_2}{\bar{V}_1}D_{21}^{n}+\frac{\bar{V}_n}{\bar{V}_1}D_{n1}^{n}\right] \quad (43a)$$

$$\widetilde{D}_{12}^{n}=D_{12}^{n}-\frac{\bar{V}_1C_1}{\bar{V}_1C_1+\bar{V}_2C_2+\bar{V}_nC_n}\left[D_{12}^{n}+\frac{\bar{V}_2}{\bar{V}_1}D_{22}^{n}+\frac{\bar{V}_n}{\bar{V}_1}D_{n2}^{n}\right] \quad (43b)$$

$$\widetilde{D}_{21}^{n}=D_{21}^{n}-\frac{\bar{V}_2C_2}{\bar{V}_1C_1+\bar{V}_2C_2+\bar{V}_nC_n}\left[\frac{\bar{V}_1}{\bar{V}_2}D_{11}^{n}+D_{21}^{n}+\frac{\bar{V}_n}{\bar{V}_2}D_{n1}^{n}\right] \quad (43c)$$

$$\widetilde{D}_{22}^{n}=D_{22}^{n}-\frac{\bar{V}_2C_2}{\bar{V}_1C_1+\bar{V}_2C_2+\bar{V}_nC_n}\left[\frac{\bar{V}_1}{\bar{V}_2}D_{12}^{n}+D_{22}^{n}+\frac{\bar{V}_n}{\bar{V}_2}D_{n2}^{n}\right] \quad (43d)$$

Therefore, a general equation can be expressed as

$$\widetilde{D}_{ij}^{n}=D_{ij}^{n}-Z_i^{PT}\sum_{k=1}^{n}\frac{\bar{V}_k}{\bar{V}_i}D_{kj}^{n} \quad (44a)$$

where $Z_{i(1,2)}^{PT}=\frac{\bar{V}_iC_i}{\bar{V}_1C_1+\bar{V}_2C_2+\bar{V}_nC_n}$ is a concentration-normalized variable in the PT diffusion couple.

Instead of a PT system, if we consider a ternary conventional diffusion couple in which all the elements (1, 2 and *n*) produce the diffusion profile, we have $\bar{V}_1C_1+\bar{V}_2C_2+\bar{V}_nC_n=1$, such that this interdiffusion and intrinsic diffusion coefficients can be expressed as

$$\widetilde{D}_{ij}^{n}=D_{ij}^{n}-C_i\sum_{k=1}^{n}\bar{V}_kD_{kj}^{n} \quad (44b)$$

The correlation between intrinsic and tracer diffusion coefficients needs to be derived now. Eq. 20 in a PT diffusion couple reduces to (since elements 3 to n-1 remain constant)

$$\frac{\partial \mu_i}{\partial x}=\frac{RT}{C_1}\varphi_{i1}^{n}\frac{\partial C_1}{\partial x}+\frac{RT}{C_{i2}}\varphi_{ii}^{n}\frac{\partial C_i}{\partial x} \quad (45)$$

Replacing this in Eq. 21, we have

$$J_i = -\frac{C_i D_i^*}{RT}\left(\frac{RT}{C_1}\ \varphi_{i1}^n \frac{\partial C_1}{\partial x} + \frac{RT}{C_{i2}}\varphi_{i2}^n \frac{\partial C_2}{\partial x}\right) - (\xi_{PT} C_i D_i^*)\frac{\overline{V}_1 C_1 D_1^*}{RT}\left(\frac{RT}{C_1}\ \varphi_{i1}^n \frac{\partial C_1}{\partial x} + \frac{RT}{C_{i2}}\varphi_{i2}^n \frac{\partial C_2}{\partial x}\right) \dots - (\xi_{PT} C_i D_i^*)\frac{\overline{V}_2 C_2 D_2^*}{RT}\left(\frac{RT}{C_1}\ \varphi_{i1}^n \frac{\partial C_1}{\partial x} + \frac{RT}{C_{i2}}\varphi_{i2}^n \frac{\partial C_2}{\partial x}\right) - (\xi_{PT} C_i D_i^*)\frac{\overline{V}_n C_n D_n^*}{RT}\left(\frac{RT}{C_1}\ \varphi_{i1}^n \frac{\partial C_1}{\partial x} + \frac{RT}{C_{i2}}\varphi_{i2}^n \frac{\partial C_2}{\partial x}\right) \tag{46}$$

Similarly, from Eq. 38, we have

$$J_i = -\sum_{j=1}^{2} D_{ij}^n \frac{\partial C_j}{\partial x} = -D_{i1}^n \frac{\partial C_1}{\partial x} - D_{i2}^n \frac{\partial C_2}{\partial x} \tag{47}$$

Comparing Eq. 46 and 47 for $\frac{\partial C_1}{\partial x}$ and $\frac{\partial C_2}{\partial x}$, we have

$$D_{ij}^n = \frac{C_i}{C_j} D_i^* \varphi_{ij}^n + \left(\xi_{PT} C_i D_i^*\right)\frac{C_1}{C_j}\overline{V}_1 D_1^* \varphi_{1j}^n + \left(\xi_{PT} C_i D_i^*\right)\frac{C_2}{C_j}\overline{V}_2 D_2^* \varphi_{2j}^n + \left(\xi_{PT} C_i D_i^*\right)\frac{C_n}{C_j}\overline{V}_n D_n^* \varphi_{nj}^n$$

$$D_{ij}^n = \frac{C_i}{C_j} D_i^* \varphi_{ij}^n \left(1 + W_{ij}^{n,a}\right) \tag{48}$$

Where the vacancy wind effect for the intrinsic diffusion coefficients is (for k = 1, 2 and n)

$$1 + W_{ij}^{PT} = 1 + \frac{\xi_{PT}}{\varphi_{ij}^n}\sum_{k=1}^{n} \overline{V}_k C_k D_k^* \varphi_{kj}^n = 1 + \frac{2\sum_{k=1}^{n} \overline{V}_k C_k D_k^* \varphi_{kj}^n}{\varphi_{ij}^n M_o \sum_{j=1}^{n} \overline{V}_j C_j D_j^*}.$$

Again, note that this correlation is the same for the ternary (T) system of elements 1, 2, and *n* such that we have $\xi_{PT} = \xi_T$. However, the correlation between interdiffusion and intrinsic diffusion coefficients is not the same for T and PT systems, which is already discussed.

## 4. Discussion estimating the diffusion coefficients

Calculating the intrinsic and interdiffusion coefficients from the tracer diffusion coefficients makes it easier to understand the role of the vacancy wind effect. As mentioned already, the analysis in ternary and multicomponent systems was invariably done by considering constant molar volume since the actual molar volume variation is unknown. This may vary either ideally (linearly) following Vegard's law (rule of mixture) or non-ideally, in which the molar volume variation may deviate positively or negatively from the linear variation. A recent analysis in the binary system indicated that consideration of linear variation estimates data with little error instead of consideration of constant molar volume [18]. The equation scheme established in this study can be utilized for both linear and non-linear variation of the molar volume. Since the actual molar volume variation is generally not known in a ternary and multicomponent system, the data are calculated and compared for constant and linear variation of molar volume.

The intrinsic diffusion coefficients can be calculated from Eq. 24 for constant molar volume. For ideal, i.e. linear molar volume variation, the partial molar volume of an element $k$ ($\bar{V}_k$) is equal to the molar volume of the pure element ($V_m^k$). Therefore, we can calculate this by replacing $\bar{V}_k$ with $V_m^k$ in Eq. 23. The vacancy wind effect depends on the difference in diffusivities of the elements and also the composition at which these are estimated. An alloy rich in element 1 will have a higher vacancy wind effect for this element compared to the elements present in low concentration. To avoid this issue, the differences in vacancy wind effect on different elements are studied only because of differences in diffusivities and molar volumes. The analysis in this study is done on the equiatomic composition.

Let us first consider the NiCoFeCr system in which the tracer diffusion coefficients of all the elements estimated following the radiotracer method are available at the equiatomic composition [31]. At 1100ºC, the values are reported as $D_1^*(D_{Cr}^*) = 2\times10^{-15}$, $D_2^*(D_{Fe}^*) = 1.1\times10^{-15}$, $D_3^*(D_{Co}^*) = 0.45\times10^{-15}$, $D_4^*(D_{Ni}^*) = 0.46 \times 10^{-15}\ m^2/s$. The molar volume of the elements is calculated from the lattice parameter data [32] as $V_m^1(V_m^{Cr}) = 7.22 \times 10^{-6}, V_m^2(V_m^{Fe}) = 6.85 \times 10^{-6}$, $V_m^3(V_m^{Co}) = 6.67 \times 10^{-6}$, $V_m^4\left(V_m^{Ni}\right) = 6.57 \times 10^{-6}\ m^3/mol$. The differences in the molar volume and diffusivities of the elements in this system are relatively low. Therefore, after analyzing this system, we shall consider another system with a higher possible difference in molar volume and diffusivities to compare the data for these two different cases. The thermodynamic factor for constant molar volume variation, i.e. $\emptyset_{ij}^n$ (by taking the gradient of $\partial lna_i$ vs. $\partial lnN_i$ plot) and considering ideal molar volume variation, i.e. $\varphi_{ij}^n = \frac{\partial lna_i}{\partial lnC_j} - \frac{V_jC_j}{V_nC_n}\frac{\partial lna_i}{\partial lnC_n}$ (by taking the gradient of $\partial lna_i$ vs. $\partial lnC_i$ plot) at the equiatomic composition are calculated, as listed in Table 1. The activity data for this calculation is extracted from the TCHE3 database of ThermoCalc for the NiCoFeCr system at 1100ºC [33]. The intrinsic diffusion coefficients for four different conditions are calculated for different conditions, i.e. (i) constant molar volume, i.e. neglecting the molar volume variation, and vacancy wind effect (N-V, N-W), (ii) neglecting the molar volume variation but considering vacancy wind effect (N-V, C-W), (iii) considering the molar volume variation but neglecting the vacancy wind effect (C-V, N-W), and (iv) considering molar volume variation and the vacancy wind effect (C-V, C-W). These are listed in Table 2. The contribution of the vacancy wind effect $(1 + W_{ij})$ can be understood by the deviation of these values from 1. The higher the deviation indicates, the higher the vacancy wind effect, i.e., the role of cross terms of the Onsager formalism. It can be noticed that certain cross-intrinsic diffusion (which are not negligible compared to the main intrinsic diffusion coefficients) have a significant role in the vacancy wind effect, for example, $D_{12}^{4}\left(D_{CrFe}^{Ni}\right)$, $D_{21}^{4}\left(D_{FeCr}^{Ni}\right)$, and $D_{41}^{4}\left(D_{NiCr}^{Ni}\right)$. As already mentioned, the diffusion coefficients in ternary and multicomponent systems are generally calculated until now, considering the constant molar volume and neglecting the vacancy wind effect (N-V, N-W). On the other hand, this study proposes the calculation by considering the molar volume of diffusing elements and the vacancy wind effect (C-V, C-W). The comparisons between these two data sets and the percentage differences are shown in Fig. 1. Although there is not much difference in diffusivities and molar volumes of the elements in this system, the differences in data for certain diffusivities cannot be neglected. The effect needs to be judged for the ratio of diffusivities. Relative mobilities (i.e. ratio of tracer diffusion coefficients) or ratio of main intrinsic diffusion coefficients indicate the influence of elements on rate-controlling processes. The ratio of main intrinsic diffusion coefficients is compared with the ratio of tracer diffusion coefficients to indicate the combined effect of molar volume and the vacancy wind effects, as shown in Fig. 2. The ratio of main intrinsic diffusion coefficients of elements 1 (Cr) and 2 (Fe), i.e. $\frac{D_{11}^{4}\left(D_{CrCr}^{Ni}\right)}{D_{22}^{4}\left(D_{FeFe}^{Ni}\right)}$, is increased by 20.9% from the ratio of tracer diffusion coefficients, i.e. $\frac{D_1^*(D_{Cr}^*)}{D_2^*(D_{Fe}^*)}$ when we neglect molar volume variation and the vacancy wind effect. However, this is increased by 25.3% when we consider both the elements' molar volumes and the vacancy wind effect from the ratio of tracer diffusion coefficients. On the other hand, the ratio of the main intrinsic diffusion coefficients of elements 1 (Cr) and 3 (Co), i.e. $\frac{D_{11}^{4}\left(D_{CrCr}^{Ni}\right)}{D_{33}^{4}\left(D_{CoCo}^{Ni}\right)}$, is increased by 31.5% when we neglect molar volume variation and vacancy wind effect from the ratio of tracer diffusion coefficients $\frac{D_1^*(D_{Cr}^*)}{D_3^*(D_{Co}^*)}$. It is increased by 48.4% when considering both the elements' molar volumes and the vacancy wind effect. Therefore, the higher the difference between tracer diffusion coefficients and molar volume, the higher the error (and not negligible) in calculation considering the constant molar volume and neglecting the vacancy wind effect.

Next, we calculate the interdiffusion coefficients from the intrinsic diffusion coefficients already calculated. These can be calculated considering the molar volume of diffusing elements following Eq. 10b and considering the constant molar volume following Eq. 10c. These are listed in Table 3. It can be seen that the contributions of the vacancy wind effect on interdiffusion coefficients are far less compared to the vacancy wind effect on the intrinsic diffusion coefficients. However, it should be noted that the interdiffusion coefficients are the average of *n* intrinsic diffusion coefficients (refer to Eq. 10) and do not reflect on the actual diffusion coefficients of elements. Although it is essential to correlate these diffusion parameters with the diffusion profiles, they are vague [34] and can be very often misleading depending on the composition at which these are estimated, especially in a multicomponent system. When calculated in an alloy with a low concentration of alloying elements, the interdiffusion coefficients of minor elements are almost equal to the intrinsic diffusion coefficients. However, these may be very different when estimated in an alloy with a significant concentration of alloying elements, like the equiatomic composition at which diffusion coefficients are calculated in this study. For example, intrinsic diffusion coefficients $D_{31}^{4}(D_{CoCr}^{Ni})$ and $D_{32}^{4}(D_{CoFe}^{Ni})$ considering the molar volume and vacancy wind effect (C-V, C-W) are equal to only $-0.076 \times 10^{-15}\ m^2/s$ (equal for both). However, the interdiffusion coefficients $\widetilde{D}_{31}^{4}(\widetilde{D}_{CoCr}^{Ni})$ and $\widetilde{D}_{32}^{4}(\widetilde{D}_{CoFe}^{Ni})$ are calculated as $-0.815 \times 10^{-15}$ and $-0.421 \times 10^{-15}\ m^2/s$, respectively. Since the main interdiffusion coefficient of this element is $\widetilde{D}_{33}^{4}\left(\widetilde{D}_{CoCo}^{Ni}\right) = 0.504 \times 10^{-15}\ m^2/s$, one may wrongly indicate from these cross interdiffusion coefficients that the diffusional interactions of this element with elements 1 (Cr) and 2 (Fe) are significantly high such the flux will increase significantly when element 3 (Co) diffuses in opposite (with a different sign of gradient) to element 1 (Cr) and 2 (Fe) or reduce when diffuse in the same direction (with same sign of gradient). This is incorrect since the actual diffusional interactions are reflected with the cross-intrinsic diffusion coefficients, which are much smaller than the main intrinsic diffusion coefficient $D_{33}^{4}\left(D_{CoCo}^{Ni}\right) = 0.454 \times 10^{-15}\ m^2/s$. Since interdiffusion coefficients are the average of four intrinsic diffusion coefficients in a quaternary system, this may reflect misleading diffusional interactions or relative diffusion coefficients.

As already mentioned, the difference in mobilities and molar volumes of the elements are relatively low in the NiCoFeCr system discussed above. The diffusivity of certain elements, such as Ta, Al, Cu, Si etc., is higher than Ni's diffusion rate in Ni-based alloys. Whereas W, Re, Ru, and Pt have much lower diffusivities. The difference between the fastest and slowest diffusion rates of certain elements can easily be more than one order of magnitude. Moreover, the molar volumes of Ni, Co, etc., are significantly different compared to the molar volumes of Al, W, Re, Pt, Ta etc. Therefore, one may calculate data with a substantially higher error when intrinsic and interdiffusion coefficients are calculated considering the constant molar volume instead of the elements' molar volumes. Again, such comparison is suitable for equiatomic composition to avoid composition-dependent effects on the vacancy wind effect and also for the sake of comparison with the data estimated for NiCoFeCr equiatomic composition. For this reason, we considering an imaginary system with diffusion coefficients of $D_1^* = 10\times10^{-15}$, $D_2^* = 5\times10^{-15}$, $D_3^* = 2\times10^{-15}$, $D_4^* = 1 \times 10^{-15}\ m^2/s$ and molar volumes of these elements as $V_m^1 = 9 \times 10^{-6}, V_m^2 = 8 \times 10^{-6}, V_m^3 = 7 \times 10^{-6}, V_m^4 = 6 \times 10^{-6}\ m^3/mol$. Moreover, for the sake of comparison of data without the effect of activity data, we assume the same activity data as considered for NiCoFeCr system at 1100 ºC. Therefore, the thermodynamic factors calculated considering the constant molar volume ($\emptyset_{ij}^4$) remains the same as listed in Table 1a, which are also listed in Table 4a for comparison. However, the thermodynamic factors calculated considering the molar volume variation ($\varphi_{ij}^4$) variation are different, as shown in Table 4b. The data calculated for this are higher because of higher molar volume differences than those calculated for the NiCoFeCr system, as shown in Table 1b.

Following, again the intrinsic diffusion coefficients are calculated for four different conditions (i) constant molar volume i.e. neglecting the molar volume variation, and vacancy wind effect (N-V, N-W), (ii) neglecting the molar volume variation but considering vacancy wind effect (N-V, C-W), (iii) considering the molar volume variation but neglecting the vacancy wind effect (C-V, N-W), and (iv) considering molar volume variation and the vacancy wind effect (C-V, C-W) considering Eq. 23 and 24. The estimated data are listed in Table 5, indicating higher differences in data, which could be understood from the plots shown in Fig. 3 in comparison to data plotted for NiCoFeCr system in Fig. 1. This is also evident from the ratios of main intrinsic diffusion coefficients compared to the ratios of tracer diffusion coefficients, as shown in Fig. 4 in comparison to the data plotted for the NiCoFeCr system in Fig. 2. Following the interdiffusion coefficients are calculated and compared in Table 6. It can be seen again that the interdiffusion coefficients may indicate very wrong diffusional interactions if we discuss considering the interdiffusion coefficients instead of the intrinsic diffusion coefficients. For example, cross interdiffusion coefficients $\widetilde{D}_{21}^{4}$, $\widetilde{D}_{31}^{4}$ and $\widetilde{D}_{32}^{4}$ has much higher value compared to the cross intrinsic diffusion coefficients $D_{21}^{4}$, $D_{31}^{4}$ and $D_{31}^{4}$. Therefore, again, we can conclude that the discussion of diffusional interactions with cross-interdiffusion coefficients could be very wrong instead of discussion considering the cross intrinsic diffusion coefficients.

**5. Conclusion**

The correlation between tracer and intrinsic diffusion coefficients are derived in this study for ternary and multicomponent systems considering the molar volume of diffusing elements and Onsager's cross phenomenological constants (vacancy wind effect). These effects are then discussed by calculating intrinsic and interdiffusion coefficients in quaternary systems. The following conclusions are drawn from this study:

(i) The estimation of diffusion coefficients considering the molar volumes of the diffusing elements is very important for more accurate estimation.

(ii) The diffusion coefficients should be calculated considering the vacancy wind effect, which may strongly influence depending on the compositions at which these are estimated.

(iii) Until now the diffusion coefficients in ternary and multicomponent systems are mostly calculated considering a constant molar volume and neglecting the contribution of vacancy wind effect. This study indicates a significant difference in estimated data when we consider the molar volumes of the diffusing elements and the vacancy wind effect.

(iv) Since intrinsic diffusion coefficients are rarely estimated in ternary and multicomponent systems, discussing the diffusional interactions between elements with interdiffusion coefficients is a common practice. However, the data estimated in this study clearly indicates that such a discussion considering interdiffusion coefficients instead of intrinsic diffusion coefficients could be misleading.

(v) The equation scheme established and analysis conducted indicate the need for a major change in estimation methodology in ternary and multicomponent systems.

**Acknowledgement:** The author acknowledges the help from Suman Sadhu in extracting the thermodynamic details required for this analysis.

| TF (constant $V_m$) | | |
|---|---|---|
| $\phi^4_{11}$ | $\phi^4_{12}$ | $\phi^4_{13}$ |
| 1.334 | 0.198 | -0.074 |
| $\phi^4_{21}$ | $\phi^4_{22}$ | $\phi^4_{23}$ |
| 0.124 | 1.103 | -0.054 |
| $\phi^4_{31}$ | $\phi^4_{32}$ | $\phi^4_{33}$ |
| -0.326 | -0.238 | 1.016 |
| $\phi^4_{41}$ | $\phi^4_{42}$ | $\phi^4_{43}$ |
| -1.131 | -1.059 | -0.876 |

(a)

| TF (Actual $V_m$) | | |
|---|---|---|
| $\varphi^4_{11}$ | $\varphi^4_{12}$ | $\varphi^4_{13}$ |
| 1.336 | 0.199 | -0.074 |
| $\varphi^{Ni}_{21}$ | $\varphi^{Ni}_{22}$ | $\varphi^{Ni}_{23}$ |
| 0.127 | 1.115 | -0.054 |
| $\varphi^{Ni}_{31}$ | $\varphi^{Ni}_{32}$ | $\varphi^{Ni}_{33}$ |
| -0.334 | -0.241 | 1.020 |
| $\varphi^4_{41}$ | $\varphi^4_{42}$ | $\varphi^4_{43}$ |
| -1.159 | -1.070 | -0.880 |

(b)

Table 1 Thermodynamic factors (TF) calculated at the equiatomic composition of NiCoFeCr at 1100 °C by extracting the activity data from the thermodynamic database TCHEA3 [33] for (a) constant molar volume $\emptyset^n_{ij} = \frac{\partial ln a_i}{\partial ln N_j}$ and (b) considering the molar volumes of diffusing elements assuming ideal variation $\varphi^n_{ij} = \frac{\partial ln a_i}{\partial ln C_j}$ (refer 1=Cr, 2=Fe, 3=Co and 4=Ni). Molar volumes of the elements. $V^1_m = 7.22×10^{-6}, V^2_m = 6.85×10^{-6}, V^3_m = 6.67×10^{-6}, V^4_m = 6.57×10^{-6}\ m^3/mol.$

| $D^n_{ij}$ ($\times 10^{-15}\ m^2/s$) | Constant $V_m$ | | | Actual $V_m$ | | |
|---|---|---|---|---|---|---|
| | N-V, N-W | N-V, C-W | (1+$W_{ij}$) | C-V, N-W | C-V, C-W | (1+$W_{ij}$) |
| $D^4_{11}$ | 2.668 | 2.966 | 1.11 | 2.672 | 2.993 | 1.12 |
| $D^4_{12}$ | 0.396 | 0.538 | 1.36 | 0.398 | 0.545 | 1.37 |
| $D^4_{13}$ | -0.148 | -0.169 | 1.14 | -0.148 | -0.169 | 1.14 |
| $D^4_{21}$ | 0.136 | 0.300 | 2.21 | 0.134 | 0.306 | 2.28 |
| $D^4_{22}$ | 1.213 | 1.291 | 1.06 | 1.227 | 1.313 | 1.07 |
| $D^4_{23}$ | -0.059 | -0.071 | 1.20 | -0.059 | -0.071 | 1.20 |
| $D^4_{31}$ | -0.147 | -0.080 | 0.54 | -0.150 | -0.076 | 0.51 |
| $D^4_{32}$ | -0.107 | -0.075 | 0.70 | -0.108 | -0.076 | 0.70 |
| $D^4_{33}$ | 0.457 | 0.452 | 0.99 | 0.459 | 0.454 | 0.99 |
| $D^4_{41}$ | -0.520 | -0.452 | 0.87 | -0.533 | -0.458 | 0.86 |
| $D^4_{42}$ | -0.487 | -0.455 | 0.93 | -0.492 | -0.458 | 0.93 |
| $D^4_{43}$ | -0.403 | -0.408 | 1.01 | -0.405 | -0.409 | 1.01 |

Table 2 Intrinsic diffusion coefficients ($D^n_{ij}$) calculated for constant molar volume and actual molar volume neglecting the vacancy wind effect (N-VW) and considering the vacancy wind effect (C-VW) at equiatomic composition at 1100 °C The tracer diffusion coefficients required for these calculations are taken from the Ref. [31] as $D^*_1 = 2×10^{-15}$,$D^*_2 = 1.1×10^{-15}$,$D^*_3 = 0.45×10^{-15}$,$D^*_4 = 0.46× 10^{-15}\ m^2/s$ (refer 1=Cr, 2=Fe, 3=Co and 4=Ni).

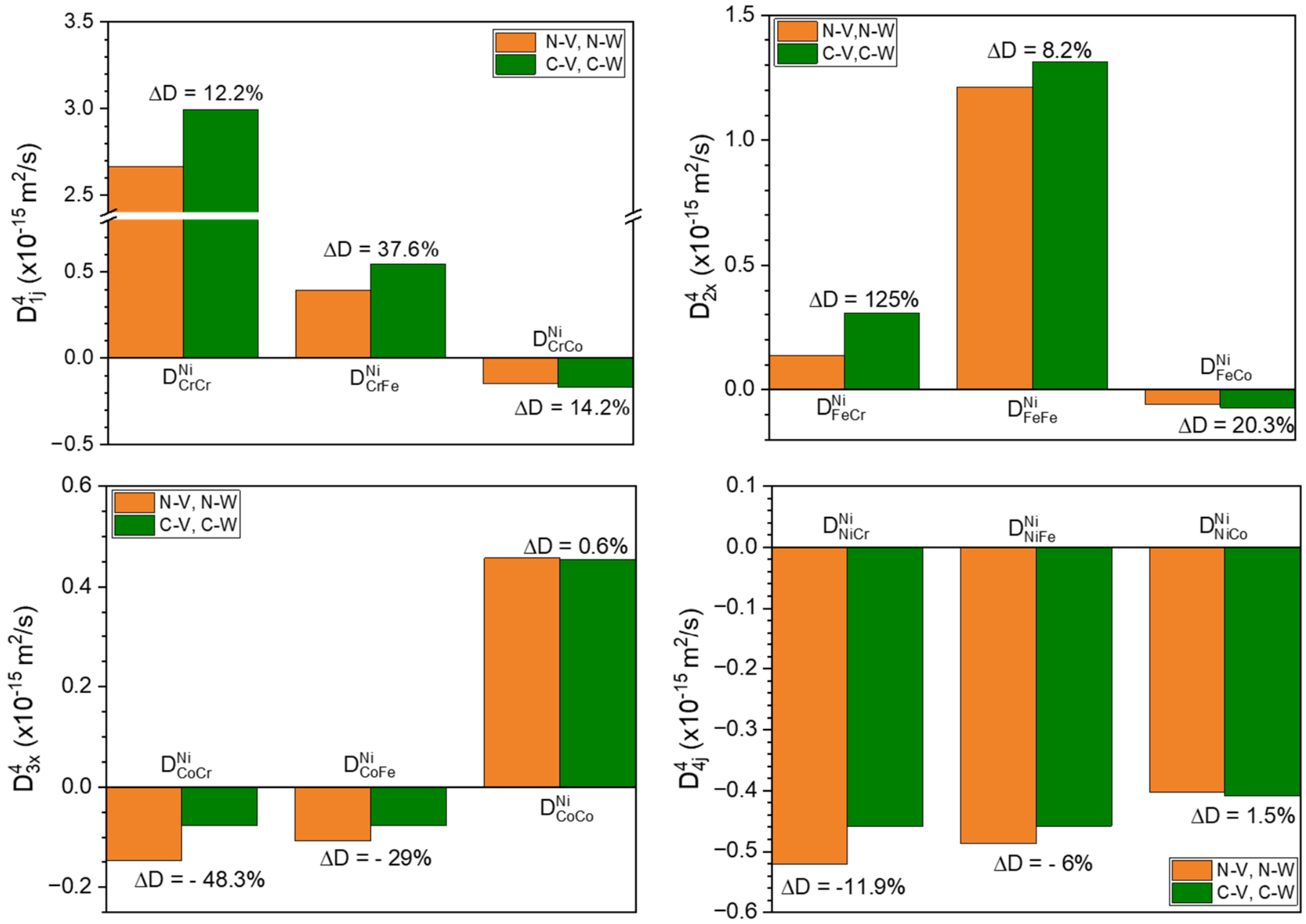

Fig. 1 Intrinsic diffusion coefficients for (a) $D^n_{1j}$, (b) $D^n_{2j}$, (c) $D^n_{3j}$, (d) $D^n_{4j}$ are compared in NiCoFeCr system (refer 1=Cr, 2=Fe, 3=Co and 4=Ni) between the data calculated by neglecting the molar volume and vacancy wind effect (N-V, N-W) and considering the molar volume and vacancy wind effect (C-V, C-W).

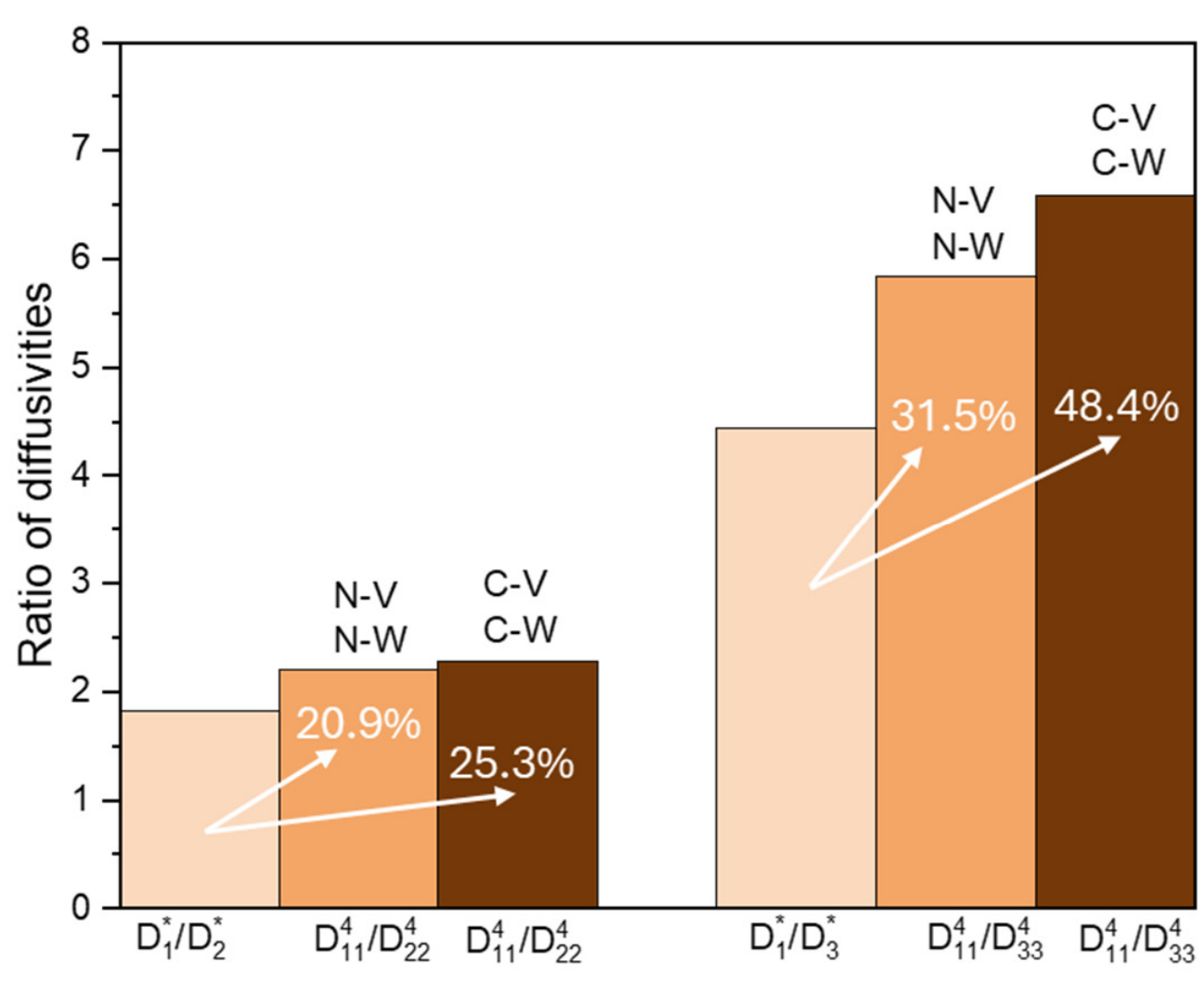

Fig. 2 Increase in ratio of main intrinsic diffusion coefficients from the ratio of tracer diffusion coefficients at the equiatomic composition of in the NiCoFeCr (refer 1=Cr, 2=Fe, 3=Co and 4=Ni) for the data calculated by neglecting the molar volume and vacancy wind effect (N-V, N-W) and considering the molar volume and vacancy wind effect (C-V, C-W).

| $\widetilde{D}_{ij}^{n}$ ($\times 10^{-15}\ m^2/s$) | Constant $V_m$ | | | Actual $V_m$ | | |
|---|---|---|---|---|---|---|
| | N-V, N-W | N-V,C-W | (1+$\widetilde{W}_{ij}$) | C-V,N-W | C-V,C-W | (1+$\widetilde{W}_{ij}$) |
| $\widetilde{D}_{11}^{4}$ | 2.134 | 2.283 | 1.07 | 2.097 | 2.254 | 1.07 |
| $\widetilde{D}_{12}^{4}$ | 0.142 | 0.213 | 1.50 | 0.130 | 0.203 | 1.56 |
| $\widetilde{D}_{13}^{4}$ | -0.110 | -0.120 | 1.09 | -0.109 | -0.119 | 1.09 |
| $\widetilde{D}_{21}^{4}$ | -0.398 | -0.384 | 1.12 | -0.441 | -0.433 | 0.98 |
| $\widetilde{D}_{22}^{4}$ | 0.959 | 0.966 | 1.00 | 0.959 | 0.968 | 1.00 |
| $\widetilde{D}_{23}^{4}$ | -0.020 | -0.022 | 1.10 | -0.020 | -0.021 | 1.05 |
| $\widetilde{D}_{31}^{4}$ | -0.681 | -0.764 | 0.54 | -0.725 | -0.815 | 1.12 |
| $\widetilde{D}_{32}^{4}$ | -0.361 | -0.340 | 0.94 | -0.376 | -0.421 | 1.12 |
| $\widetilde{D}_{33}^{4}$ | 0.495 | 0.501 | 1.01 | 0.498 | 0.504 | 1.01 |

Table 3 Interdiffusion coefficients calculated at 1100 °C at the equiatomic composition of NiCoFeCr system from the intrinsic diffusion coefficients listed in Table 2 by neglecting (N-VW) and considering (C-VW) the vacancy wind effect (refer 1=Cr, 2=Fe, 3=Co and 4=Ni).

| TF (constant $V_m$) | | |
|---|---|---|
| $\phi_{11}^{4}$ | $\phi_{12}^{4}$ | $\phi_{13}^{4}$ |
| 1.334 | 0.198 | -0.074 |
| $\phi_{21}^{4}$ | $\phi_{22}^{4}$ | $\phi_{23}^{4}$ |
| 0.124 | 1.103 | -0.054 |
| $\phi_{31}^{4}$ | $\phi_{32}^{4}$ | $\phi_{33}^{4}$ |
| -0.326 | -0.238 | 1.016 |
| $\phi_{41}^{4}$ | $\phi_{42}^{4}$ | $\phi_{43}^{4}$ |
| -1.131 | -1.059 | -0.876 |

(a)

| TF (Actual $V_m$) | | |
|---|---|---|
| $\varphi_{11}^{4}$ | $\varphi_{12}^{4}$ | $\varphi_{13}^{4}$ |
| 1.482 | 0.212 | -0.077 |
| $\varphi_{21}^{4}$ | $\varphi_{22}^{4}$ | $\varphi_{23}^{4}$ |
| 0.138 | 1.182 | -0.056 |
| $\varphi_{31}^{4}$ | $\varphi_{32}^{4}$ | $\varphi_{33}^{4}$ |
| -0.362 | -0.255 | 1.051 |
| $\varphi_{41}^{4}$ | $\varphi_{42}^{4}$ | $\varphi_{43}^{4}$ |
| -1.258 | -1.135 | -0.91 |

(b)

Table 4 Thermodynamic factors (TF) calculated for an imaginary quaternary system at equiatomic composition from the activity data of NiCoFeCr at 1100 °C extracted from thermodynamic database TCHEA3 [31] for (a) constant molar volume $\emptyset_{ij}^{n} = \frac{\partial ln a_i}{\partial ln N_j}$ and (b) considering the molar volumes of diffusing elements assuming ideal variation $\varphi_{ij}^{n} = \frac{\partial ln a_i}{\partial ln C_j}$. Molar volumes of the elements are assumed as $V_m^1 = 9\times10^{-6}$, $V_m^2 = 8\times10^{-6}$, $V_m^3 = 7\times10^{-6}$, $V_m^4 = 6\times10^{-6}\ m^3/mol$.

| $D_{ij}^{n}$ ($\times 10^{-15}\ m^2/s$) | Constant $V_m$ | | | Actual $V_m$ | | |
|---|---|---|---|---|---|---|
| | N-V, N-W | N-V,C-W | (1+$W_{ij}$) | C-V,N-W | C-V,C-W | (1+$W_{ij}$) |
| $D_{11}^{4}$ | 13.360 | 15.230 | 1.14 | 14.820 | 17.191 | 1.16 |
| $D_{12}^{4}$ | 1.990 | 2.906 | 1.46 | 2.120 | 3.159 | 1.49 |
| $D_{13}^{4}$ | -0.740 | -0.717 | 0.97 | -0.770 | -0.770 | 1.00 |
| $D_{21}^{4}$ | 0.635 | 1.566 | 2.47 | 0.690 | 1.870 | 2.71 |
| $D_{22}^{4}$ | 5.575 | 5.978 | 1.07 | 5.910 | 6.442 | 1.09 |
| $D_{23}^{4}$ | -0.270 | -0.259 | 0.96 | -0.280 | -0.280 | 1.00 |
| $D_{31}^{4}$ | -0.668 | -0.274 | 0.41 | -0.724 | -0.253 | 0.35 |
| $D_{32}^{4}$ | -0.482 | -0.291 | 0.60 | -0.510 | -0.301 | 0.59 |
| $D_{33}^{4}$ | 2.040 | 2.036 | 1.00 | 2.102 | 2.102 | 1.00 |
| $D_{41}^{4}$ | -1.159 | -0.942 | 0.81 | -1.258 | -1.019 | 0.81 |
| $D_{42}^{4}$ | -1.070 | -0.966 | 0.90 | -1.135 | -1.033 | 0.91 |
| $D_{43}^{4}$ | -0.876 | 0.869 | 0.99 | -0.910 | -0.910 | 1.00 |

Table 5 Intrinsic diffusion coefficients ($D_{ij}^{n}$) calculated at equiatomic composition of an imaginary quaternary system for constant molar volume and actual molar volume neglecting the vacancy wind effect (N-VW) and considering the vacancy wind effect (C-VW) at equiatomic composition. The tracer diffusion coefficients considered for these calculations as $D_1^* = 10\times10^{-15}$,$D_2^* = 5\times10^{-15}$,$D_3^* = 2\times10^{-15}$,$D_4^* = 1\times 10^{-15}\,\frac{m^2}{s}$.

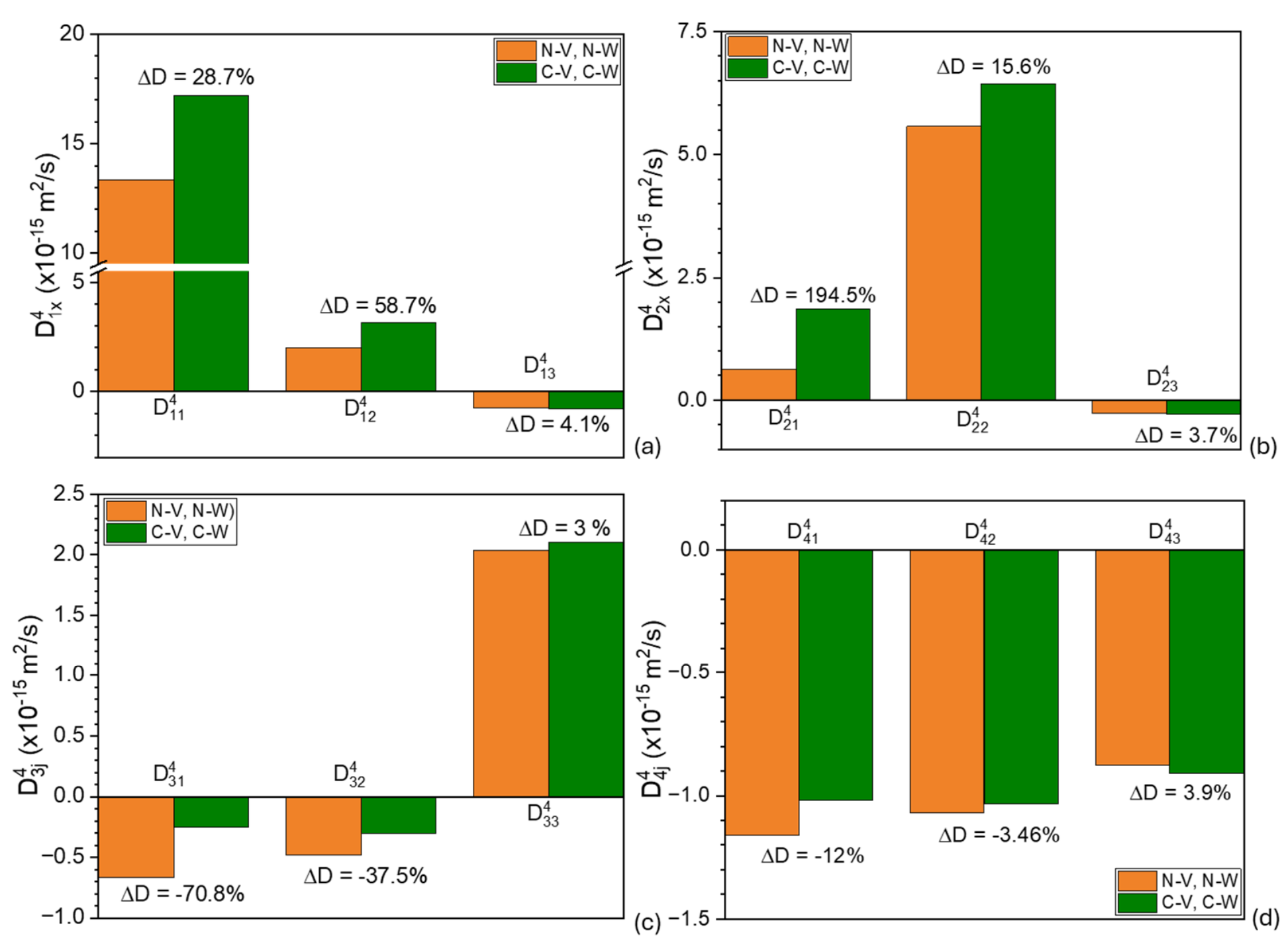

Fig. 3 Intrinsic diffusion coefficients for (a) $D_{1j}^{n}$, (b) $D_{2j}^{n}$, (c) $D_{3j}^{n}$ , (d) $D_{4j}^{n}$ in the imaginary system compared between the data calculated by neglecting the molar volume and vacancy wind effect (N-V, N-W) and considering the molar volume and vacancy wind effect (C-V, C-W).

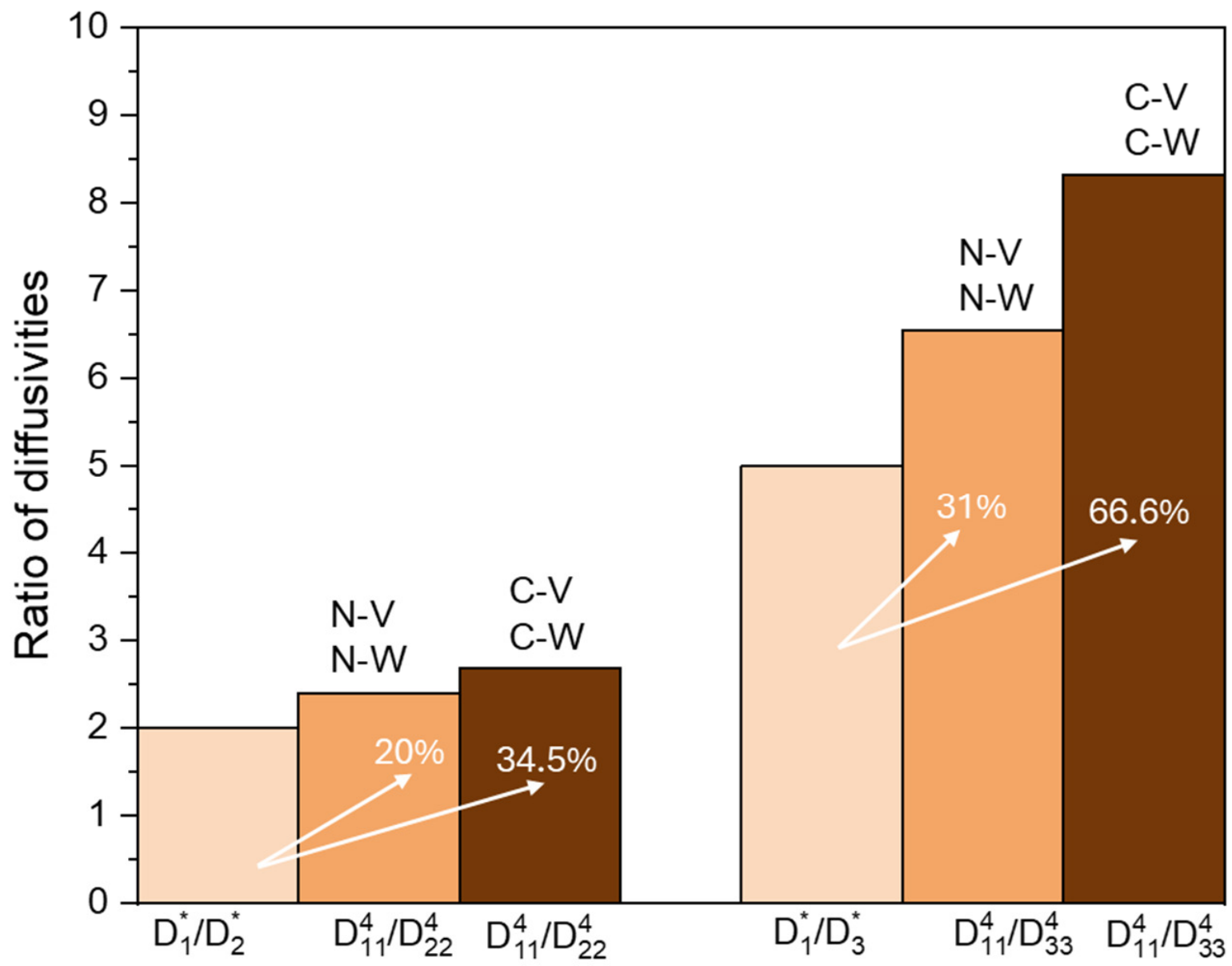

Fig. 4 Increase in ratio of main intrinsic diffusion coefficients from the ratio of tracer diffusion coefficients at the equiatomic composition of an imaginary system for the data calculated by neglecting the molar volume and vacancy wind effect (N-V, N-W) and considering the molar volume and vacancy wind effect (C-V, C-W).

| $\tilde{D}^n_{ij}$ ($\times 10^{-15}\ m^2/s$) | Constant $V_m$ | | | Actual $V_m$ | | |
|---|---|---|---|---|---|---|
| | N-V, N-W | N-V,C-W | (1+$\tilde{W}_{ij}$) | C-V,N-W | C-V,C-W | (1+$\tilde{W}_{ij}$) |
| $\tilde{D}^4_{11}$ | 10.318 | 10.337 | 1.00 | 10.301 | 11.412 | 1.11 |
| $\tilde{D}^4_{12}$ | 0.487 | 0.999 | 2.05 | 0.254 | 0.770 | 3.03 |
| $\tilde{D}^4_{13}$ | -0.779 | -0.765 | 0.98 | -0.156 | -0.154 | 0.99 |
| $\tilde{D}^4_{21}$ | -2.407 | -2.330 | 0.97 | -3.519 | -3.523 | 1.00 |
| $\tilde{D}^4_{22}$ | 4.071 | 4.071 | 1.07 | 4.044 | 4.053 | 1.00 |
| $\tilde{D}^4_{23}$ | -0.309 | -0.307 | 0.99 | -0.334 | -0.334 | 1.00 |
| $\tilde{D}^4_{31}$ | -3.710 | -4.170 | 1.12 | -4.933 | -5.646 | 1.14 |
| $\tilde{D}^4_{32}$ | -1.985 | -2.198 | 0.94 | -2.376 | -2.690 | 1.13 |
| $\tilde{D}^4_{33}$ | 2.002 | 1.988 | 1.01 | 1.488 | 1.488 | 1.00 |

Table 6 Interdiffusion coefficients calculated at the equiatomic composition of and imaginary quaternary system from the intrinsic diffusion coefficients listed in Table 5 by neglecting (N-VW) and considering (C-VW) the vacancy wind effect.